\documentclass{emulateapj}

\pdfoutput=1
\usepackage{url}
\usepackage{float}
\usepackage{natbib}
\usepackage{amsmath}
\usepackage{epstopdf}
\usepackage{graphicx}

\def\bfref{}

\begin{document}

\title{A Comprehensive Study of Relativistic Gravity using PSR B1534+12}

\author{Emmanuel Fonseca}
\affil{Department of Physics and Astronomy, The University of British Columbia,
Vancouver, BC V6T-1Z1, Canada} 
\email{efonseca@phas.ubc.ca} 

\author{Ingrid H. Stairs}
\affil{Department of Physics and Astronomy, The University of British Columbia,  
Vancouver, BC V6T-1Z1, Canada}
\email{stairs@astro.ubc.ca}

\and

\author{Stephen E. Thorsett}
\affil{Department of Physics, Willamette University, Salem, OR 97301, USA}
\email{thorsett@willamette.edu}

\begin{abstract}
We present updated analyses of pulse profiles and their arrival-times 
from PSR B1534+12, a 37.9-ms radio pulsar in orbit with another neutron 
star. A high-precision timing model is derived from twenty-two years of 
timing data, and accounts for all astrophysical processes that systematically 
affect pulse arrival-times. Five ``post-Keplerian" parameters are measured that 
represent relativistic corrections to the standard Keplerian quantities of the 
pulsar's binary orbit. These relativistic parameters are then used to test general 
relativity by comparing the measurements with their predicted values. 
We conclude that relativity theory is confirmed to within 0.17\% of 
its predictions. Furthermore, we derive the following astrophysical results 
from our timing analysis: a distance of $d_{\textrm{GR}} = 1.051 \pm 0.005$ kpc to the 
pulsar-binary system, by relating the ``excess" orbital decay to 
Galactic parameters; evidence for pulse ``jitter" in PSR B1534+12 due to  
short-term magnetospheric activity; and evolution in pulse-dispersion properties. 
As a secondary study, we also present several analyses on pulse-structure 
evolution and its connection to relativistic precession of the pulsar's spin axis. The 
precession-rate measurement yields a value of $\Omega_1^{\textrm{spin}}$ = 
$0.59^{+0.12}_{-0.08}$ $^{\circ}$/year (68\% confidence) that is consistent with 
expectations, and represents an additional test of relativistic gravity.
\end{abstract}


\section{Introduction}

Pulsars in relativistic binary systems have provided the 
most rigorous tests of gravitational theory in strong fields to date. 
High-precision timing of such an object produces a timing model  
that describes ``post-Keplerian" (PK) effects that characterize relativistic 
corrections to the standard orbital elements \citep{dd85,dd86}, as well as 
its nominal spin, astrometric, and environmental properties. Comparisons 
between measured and expected PK parameters produce tests of the 
gravitational theory in question. The ``Hulse-Taylor" pulsar \citep{ht75a} provided 
the first such positive case for general relativity, and still serves as an excellent laboratory for 
strong-field gravity \citep{wnt10}. The recent discovery of a massive pulsar in a 
highly relativistic orbit with a white dwarf yields a testing ground for tensor-scalar 
extensions of gravitational theory \citep{afw+13}. An extensive analysis of the 
``double-pulsar" system \citep{bdp+03} constrains general relativity to 
within 0.05\% of its predictions and remains the most stringent pulsar-timing test 
so far \citep{ksm+06}.

PSR B1534+12 was discovered by \citet{wol91a} to be in a highly 
inclined, 10.1-hour binary orbit with another neutron star. Follow-up 
timing studies on this pulsar \citep{sac+98,sttw02} produced a timing 
solution that yielded measurements of five PK parameters: \(\dot{P}_b\), the 
orbital decay of the binary system; \(\dot{\omega}\), the advance of periastron 
longitude; \(\gamma\), the time-averaged gravitational-redshift and time-dilation 
parameter; \(r\textrm{ and }s\), the ``range" and ``shape" of the Shapiro time delay. 
Simultaneous measurement of these parameters produced a self-consistent 
set of tests that complemented the Hulse-Taylor results by including tests based 
only on quasi-stationary, non-radiative PK parameters \citep{twdw92}.  
Additional results were derived using the fitted timing model, including a 
precise estimate of the pulsar's distance using the measured excess of orbital 
decay due to relative motion \citep{bb96}. 



\begin{deluxetable*}{ccccccccc}
\tabletypesize{\scriptsize}
\tablecaption{Timing Parameters for Each Backend and Frequency}
\tablewidth{0in}
\tablehead{
\colhead{Parameter} & \colhead{Mark III} & \colhead{Mark IV} & \colhead{Mark IV} 
& \colhead{ASP} & \colhead{ASP} & \colhead{ASP} & \colhead{ASP} & \colhead{ASP}
}
\startdata
Frequency (MHz) \dotfill & 1400 & 430 & 1400 & 424 & 428 & 432 & 436 & 1400 \\
Bandwidth (MHz) \dotfill & 40 & 5 & 5 & 64 & 64 & 64 & 64 & 64 \\
Spectral Channels\dotfill & 32 & 1\tablenotemark{a} & 1\tablenotemark{b} & 1 & 1 & 1 
& 1 & 16\tablenotemark{c} \\ 
Number of TOAs \dotfill & 1185 & 3102 & 664 & 1204 & 1197 & 1190 & 1124 & 231 \\
Dedispersion type \dotfill & Incoh. & Coh. & Coh. & Coh. & Coh. & Coh. & Coh. & Coh. \\
Integration time (s) \dotfill & 300 & 190 & 190 & 180 & 180 & 180 & 180 & 180 \\
Date span (years) \dotfill & 1990-94 & 1998-2005 & 1998-2005 & 2004-12 & 2004-12 & 
  2004-12 & 2004-12 & 2004-12 \\
 RMS residual, \(\sigma_{rms}\) & 5.31 & 4.21 & 6.73 & 4.48 & 4.34 & 4.65 & 4.93 & 8.27
\enddata
\label{tab:backend}
\tablenotetext{a}{Four sub-bands centered at 430 MHz were taken when the Mark IV 
data were originally recorded, but were averaged together to build signal strength.}
\tablenotetext{b}{Two sub-bands centered at 1400 MHz were taken when the Mark IV 
data were originally recorded, but were also averaged together to build signal strength.}
\tablenotetext{c}{The number of actual channels recorded sometimes varied due to 
computational limitations, so this value represents a typical number of channels used.}
\end{deluxetable*}

The time-averaged pulse profile of PSR B1534+12 is undergoing 
a secular change in observed radiation pattern at a rate of 1\% per year 
\citep{arz95}. Such changes can be linked to spin-orbit coupling in a 
strong gravitational field, which results in a precession of the pulsar's spin 
axis  \citep[``relativistic spin precession";][]{ds16} and an evolving view of the 
two-dimensional beam structure \citep{kra98}. \citet*{sta04} (hereafter STA04) 
developed a general technique to characterize the overall profile shape at a given 
epoch and derive a precession rate by measuring and comparing spin-precession 
and orbital-aberration effects that produce the observed shape evolution. The results of 
this study yielded a direct measurement of the precession rate that was consistent 
with the rate predicted by general relativity, albeit with considerably limited precision. 
Furthermore, the geometry of PSR B1534+12 and its binary system was derived by 
combining these results with a rotating-vector-model  \citep[RVM,][]{rc69a} analysis 
of the evolving polarization properties. PSR B1534+12 currently  remains the only 
pulsar for which special-relativistic orbital aberration is observed. {\bfref The effects of 
relativistic spin precession on pulse structure have also been observed in 
PSR B1913+16 \citep{wrt89}, the double-pulsar system \citep{bkk+08}, and 
most dramatically in PSR J1141-6545 \citep{mks+10}.}


In this work, we report on updated timing and profile-evolution analyses of 
PSR B1534+12, using data sets that collectively span 22 years since its 
discovery. Results from the analyses described below include improvements 
in tests of general relativity, an improved measurement of the pulsar's precession 
rate, and additional findings extracted from our time series. A full 
discussion of all current results is provided in Section \ref{sec:discuss}.

\section{Observations and Reduction}
\label{sec:obsred}

Data were obtained exclusively with the 305-m Arecibo Observatory in Puerto 
Rico, using two observing frequencies and three generations of pulsar signal 
processors. Basic information regarding the data and backends used 
in this analysis are presented in Table \ref{tab:backend}, while a more detailed 
account of observing information can be found in \citet{fon12}.

Part of this set of pulse profiles and times-of-arrival (TOAs) were recorded with 
the Mark III \citep{skn+92} and Mark IV \citep{sst+00} pulsar backends. The Mark 
III system employed a brute-force pulse de-dispersion algorithm by separating each 
receiver's bandpass into distinct spectral channels with a filterbank, detecting the 
signal within each channel, and shifting the pulse profile by the predicted amount of 
dispersive delay for alignment and coherent averaging. A small amount of Mark III 
data was obtained using the coherent-dedispersion ``reticon" subsystem; these data 
were used only in the polarization analysis. The Mark IV machine was an instrumental 
upgrade which employed the now-standard coherent de-dispersion technique 
\citep{hr75} that samples and filters the data stream prior to pulse detection. 
A series of digital filters applied in the frequency domain completely remove 
the predicted dispersion signatures while retaining even greater precision than 
the Mark III counterpart. See \citet{sac+98,sttw02} for more details on these 
observing systems and reduction of PSR B1534+12 data obtained with these 
two backends.

Recent data were collected with the Arecibo Signal Processor 
\citep[ASP;][]{dem07}, a further upgrade from the Mark III/IV systems that 
retains the coherent de-dispersion technique, but first decomposes the signals 
across a bandwidth of 64 MHz into a number of 4-MHz spectral channels that 
depends on the observing frequency. We used data collected with the four 
inner-most spectral channels centered on 430 MHz, and typically sixteen channels 
centered on 1400 MHz with some variability, due to limits in computer processing 
and available receiver bandpass. While the Mark IV machine used 4-bit data 
sampling in 5-MHz-bandpass observing mode and 2-bit sampling in 10-MHz-bandpass 
observing mode, ASP always used 8-bit sampling. The coherent de-dispersion filter 
is applied to the raw, channelized data, which are then folded modulo the topocentric 
pulse period within each channel and recorded to disk, preserving polarimetric information. 

Observations were generally conducted at semi-regular intervals, with typical scan 
lengths of an hour for each frequency. Several extensive ``campaign" observations 
were also conducted at 430 MHz, which consisted of several-hour observing sessions 
performed over 12 consecutive days, in order to obtain high-precision snapshots of the 
pulse profile at different times. Campaign sessions occurred during the summers of 1998, 
1999, 2000, 2001, 2003, 2005, and 2008\label{sec:camp}. We used all available data 
for the timing analysis, and only used most of the campaign profiles and several strong 
bi-monthly scans for the profile analysis. We excluded the 2008 ASP data from the 
profile-shape analysis due to weak, heavily scintillated signals recorded during this epoch, 
but used several stronger observations during this campaign for the RVM analysis (see 
Section \ref{sec:profanalysis}).

We used the template cross-correlation algorithm developed by \citet{tay92} 
for determining pulse phases, their TOAs and uncertainties 
using a standard-template profile. A standard template was derived for the Mark III 
and Mark IV backends at each frequency by averaging several hours of consecutive 
pulse profiles; ASP TOAs were derived using the Mark IV templates. We added small 
amounts of error in quadrature or as factors to the original TOA uncertainties, in order 
to compensate for apparent systematic errors in TOAs. We also ignored TOAs with 
uncertainties greater than 10 $\mu$s; only 10\% of all available TOAs -- including 
points affected by radio frequency interference -- were excised when using this cut. 

\begin{figure}[h]
  \begin{center}
    \includegraphics[scale=0.43]{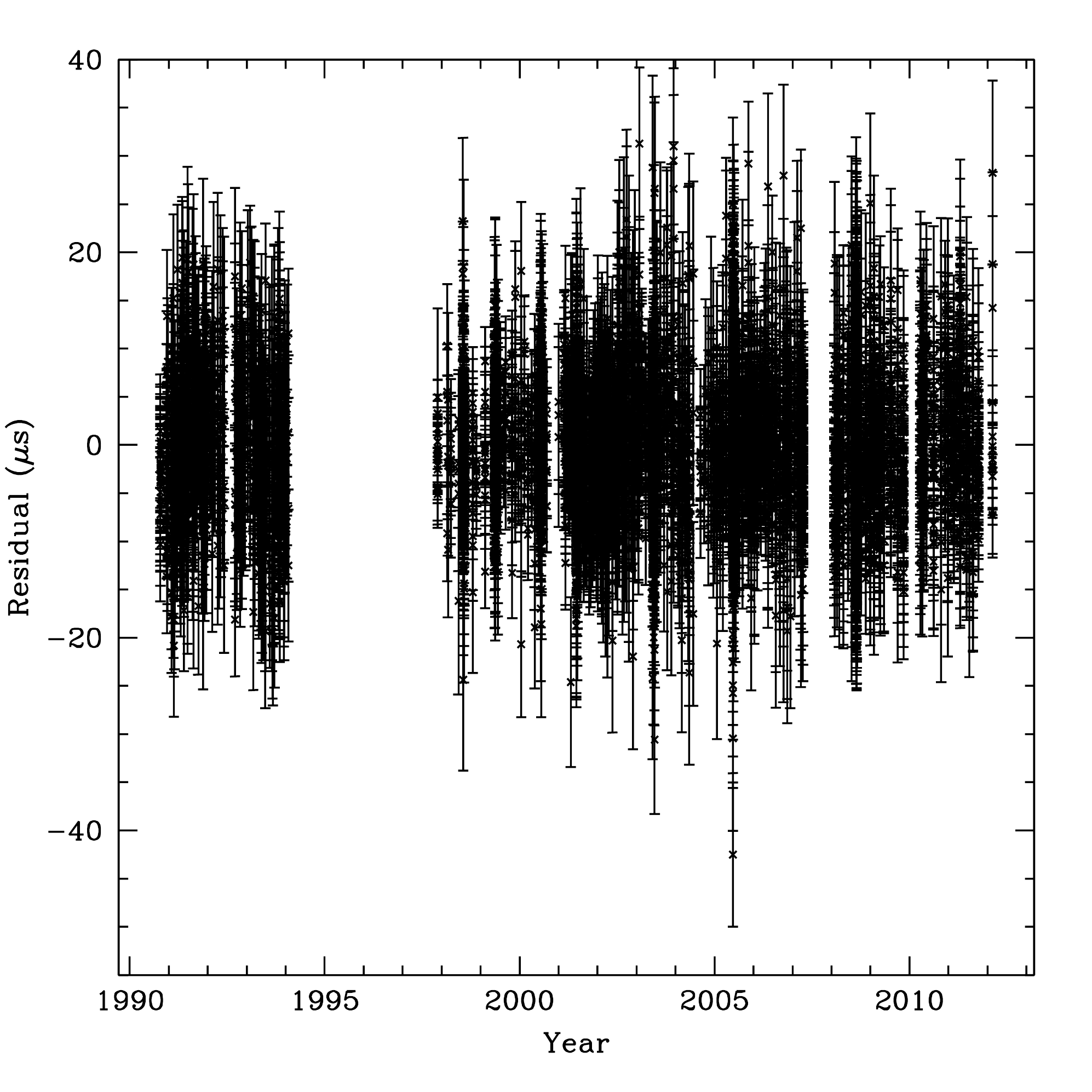}
    \caption{Post-fit timing residuals of PSR B1534+12. The best-fit RMS residual is 
    $\sigma_{\textrm{RMS}} = 4.57\textrm{ }\mu$s.}
    \label{fig:res}
  \end{center}
\end{figure}

It is important to note that there is a overlap in pulse TOAs collected with 
the Mark IV and ASP data sets between MJD 53358 and 53601. We 
incorporated TOAs acquired from both machines during this era, despite 
the overlap, due to the substantially larger ASP bandwidth; we believe 
that this difference in bandwidth does not produce many redundant data 
points. The improvement in bit sampling between backends has a measurable 
effect on the pulse profile shape, as discussed in Section \ref{sec:profanalysis} 
below.

\section{Timing Analysis} 
\label{sec:timing}

Each pulse TOA was initially recorded at a local, topocentric time $t$ and 
subsequently transformed to the Solar-system barycentric reference frame 
by accounting for a series of physical timing delays. We used the standard 
pulsar-analysis  procedure where observed TOAs and their pulse phases 
are compared to the model introduced through these timing delays by using 
a $\chi^2$-minimization fitting algorithm \citep{lk05}. {\bfref This best-fit timing solution 
was derived using the TEMPO pulsar-timing software package\footnote{ 
\url{http://tempo.sourceforge.net}}, along with the JPL DE421 planetary 
ephemeris.\footnote{\url{http://naif.jpl.nasa.gov/naif/}} The residuals between 
measured and fitted TOAs are shown in Figure \ref{fig:res}, while fitted spin, 
astrometric, and DM parameters are shown in Table \ref{tab:par1} with respect 
to the quoted reference epoch.}

The increased timespan and updated planetary ephemeris permitted 
the significant measurement of a second and third time-derivative in 
spin frequency. Due to overlap in data between the Mark IV and ASP 
machines, we first fitted for an offset between these two multi-frequency 
sets using only 430-MHz data during this timespan while holding all model 
parameters fixed. We then held this offset fixed while fitting for an additional 
offset between the Mark III and ``combined" Mark IV and ASP data during 
the global fit. 

\begin{figure}[h]
  \begin{center}
    \includegraphics[scale=0.43]{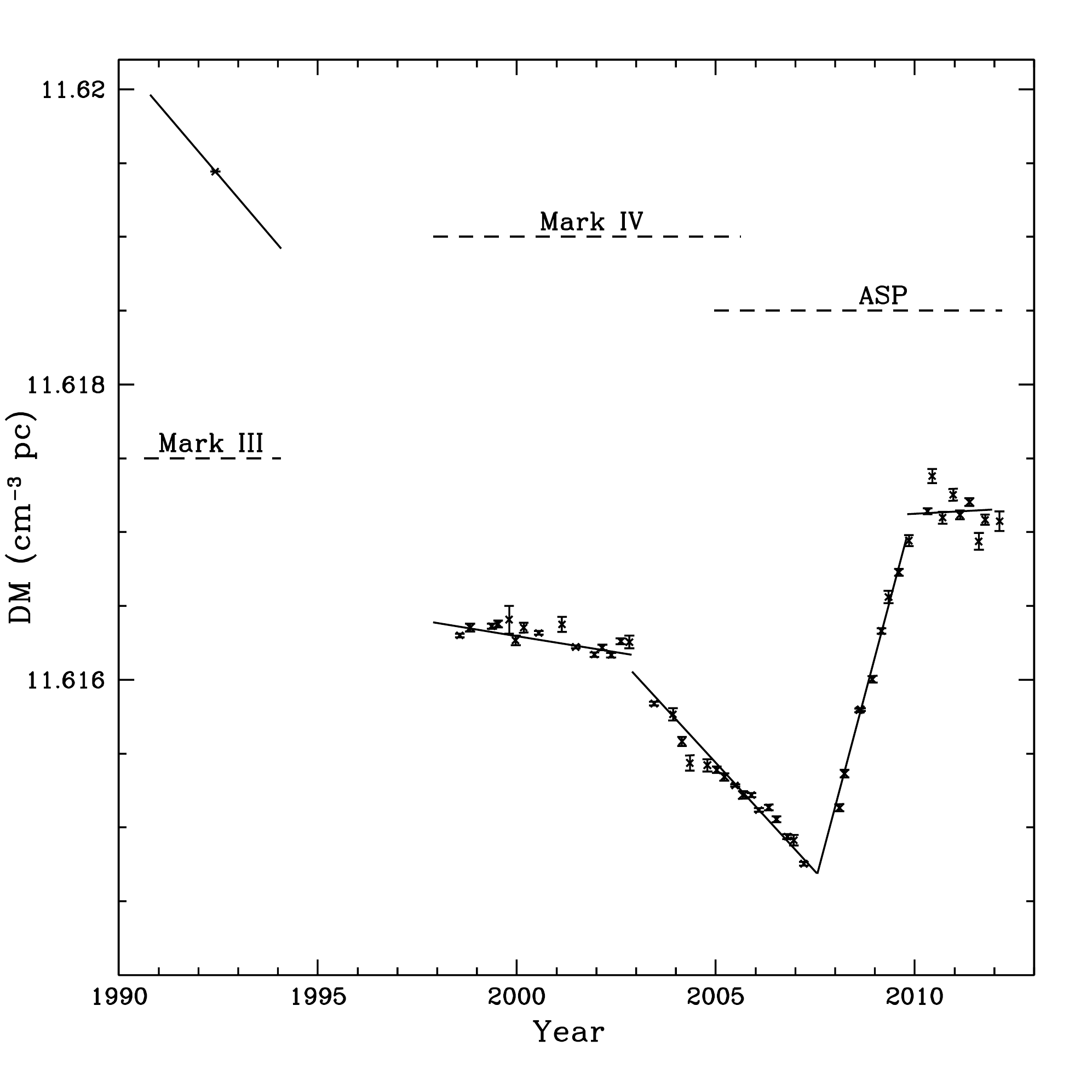}
    \caption{Dispersion measure (DM) of PSR B1534+12 versus time. The solid lines 
      represent time-derivatives that were fixed or fitted when generating 
      the global timing solution (see text). Data points and their error bars demonstrate DM 
      fitting over smaller intervals in time.}
    \label{fig:dm}
  \end{center}
\end{figure}

Moreover, we modeled the pulsar's DM in five contiguous bins due to 
observed evolution in electronic content along the line of sight to PSR 
B1534+12. An offset from a nominal DM value and a time-derivative were 
determined within each bin during the global fit. We used and fixed the 
Mark III DM bin derived by \citet{sac+98} due to systematic errors they 
found in the Mark III 430 MHz TOAs, which we discarded from this study. 
The results of this DM fitting are displayed in Figure \ref{fig:dm}. The data 
points and their error bars in Figure \ref{fig:dm} were determined by fixing 
all newly-determined parameters and fitting for DM in smaller bins of 80 
days in width, without time-derivatives. The timestamp of each small-bin 
point was computed to be the mean of all TOAs within the bin. We ultimately 
used the large-bin method during the global fit in order to minimize the total 
number of fitted parameters.

In order to extract relativistic information about this system, we first used 
the ``Damour-Deruelle" (DD) binary prescription \citep{dd86} that measures 
PK parameters in a phenomenological manner during the global fit. We 
then used the ``DDGR" model,  which assumes that general relativity is 
correct and relates each PK parameter to one or both of the binary-component 
masses \citep[e.g.][]{sac+98}, in order to determine the masses. The fitted 
DD and DDGR parameters are shown in Table \ref{tab:binarypar}.

\begin{deluxetable}{lc}
\tabletypesize{\scriptsize}
\tablecaption{Fitted Astrometric, Spin and DM Parameters}
\tablewidth{0in}
\tablehead{\colhead{Parameter} & \colhead{Value}}
\startdata
  Right Ascension, $\alpha_{\textrm{J2000}}$ \dotfill & \(15^\textrm{h}37^
     \textrm{m}09^\textrm{s}.961730(3)\) \\
  Declination, $\delta_{\textrm{J2000}}$ \dotfill & \(11^{\circ}55'55''.43387(6)\) \\
  Proper motion in R.A., \(\mu_{\alpha}\) (mas yr\(^{-1}\)) \dotfill & 1.482(7) \\
  Proper motion in Decl., \(\mu_{\delta}\) (mas yr\(^{-1}\)) \dotfill & -25.285(12) \\
  Timing parallax, \(\pi\) (mas) \dotfill & 0.86(18) \\ 
  Parameter reference epoch (MJD) \dotfill & 52077 \\
  & \\
  Rotational frequency, \(\nu\) (Hz) \dotfill & 26.38213277689397(11) \\
  First frequency derivative, \(\dot{\nu}\) (\(10^{-15} \textrm{ Hz}^2\)) \dotfill & -1.686097(2) \\
  Second freq. derivative, \(\ddot{\nu}\) (\(10^{-29} \textrm{ Hz}^3\)) \dotfill & 1.70(11) \\
  Third freq. derivative, \(\dddot{\nu}\) (\(10^{-36} \textrm{ Hz}^4\)) \dotfill & -1.6(2) \\
  & \\
  Dispersion measure, DM 1 (cm\(^{-3}\) pc) \dotfill & 11.61944(2)\tablenotemark{a} \\
  DM derivative 1 (cm\(^{-3}\) pc yr\(^{-1}\)) \dotfill & -0.000316(10)\tablenotemark{a} \\ 
  Bin 1 range, epoch (MJD) \dotfill & 48178-49380, 48778\tablenotemark{a} \\
  & \\
  DM 2 (cm\(^{-3}\) pc) \dotfill & 11.616279(14) \\ 
  DM 2 derivative (cm\(^{-3}\) pc yr\(^{-1}\)) \dotfill & -0.000043(8) \\
  Bin 2 range, epoch (MJD) \dotfill & 50775-52600, 51687.5 \\
  & \\ 
  DM 3 (cm\(^{-3}\) pc) \dotfill & 11.61537(2) \\
  DM 3 derivative (cm\(^{-3}\) pc yr\(^{-1}\)) \dotfill & -0.000294(7) \\
  Bin 3 range, epoch (MJD) \dotfill & 52601-54300, 53450.5 \\
  & \\
  DM 4 (cm\(^{-3}\) pc) \dotfill & 11.61583(8) \\
  DM 4 derivative (cm\(^{-3}\) pc yr\(^{-1}\)) \dotfill & 0.00101(3) \\ 
  Bin 4 range, epoch (MJD) \dotfill & 54301-55125, 54713 \\
  & \\
  DM 5 (cm\(^{-3}\) pc) \dotfill & 11.61713(10) \\
  DM derivative 5 (cm\(^{-3}\) pc yr\(^{-1}\)) \dotfill & -0.00001(5) \\
  Bin 5 range, epoch (MJD) \dotfill & 55126-55974, 55550 \\
\enddata
\label{tab:par1}
\tablenotetext{a}{Taken from \citet{sac+98} and fixed during the global fit.}
\tablecomments{Values in parentheses denote the uncertainty in the preceding 
digit(s).}
\end{deluxetable}

{\bfref All reported uncertainties for fitted timing parameters are the 1-$\sigma$ 
TEMPO values determined from the global fit. It is common practice in pulsar 
timing analyses to report uncertainties that are twice as large as the formal 
fitted errors, which was done in previous timing studies of PSR B1534+12. 
However, we confirmed the TEMPO 68-\% confidence intervals reported 
in Tables \ref{tab:par1} and \ref{tab:binarypar} using a Markov Chain Monte Carlo 
\citep[MCMC, e.g.][]{gre05bayes} simulation of the TEMPO global fit, assuming 
that the joint posterior distribution of the fit is a normal distribution in the $\chi^2$ 
statistic for our timing model (see Zhu et al., in preparation). Because of the 
more limited analysis of measurement uncertainties in earlier studies, we 
believe that previously published error bars have been modestly but systematically 
conservative. The substantial improvement in our measurements is 
therefore attributable both to our significantly larger data set and our improved 
uncertainty analysis.}


\section{Profile-Evolution Analysis}
\label{sec:profanalysis}


An observed pulse produces a set of Stokes-vector pulse profiles, from which a 
TOA can be derived when cross-correlating the total-intensity profile $P$ with a 
standard template as described in Sections \ref{sec:obsred}. As with pulsar timing, 
strong-field effects can give rise to observable changes in pulse-structure 
parameters, such as pulse-profile shape and polarization properties, over a 
variety of timescales. In order to detect such changes, we shifted our 430-MHz profiles 
to a common phase using the derived DD-binary timing model described in Section 
\ref{sec:timing}. Each set of campaign data was then binned into twelve 
orbital-phase cumulative profiles, while several strong bi-monthly scans were 
averaged into single profiles recorded at their respective epochs. We subsequently 
performed two distinct analyses on these total-intensity and polarization data 
in order to extract gravitational information from independent 
techniques, as described below.

For the first analysis, we employed the model developed by STA04 that establishes 
pulse-structure data as functions of time and location within the relativistic orbit. 
Values of the total-intensity profile shape $F$ at a given epoch were derived by first 
applying a principal component analysis (PCA) on a set of total-intensity profiles 
collected over time; the first and second principal components correspond to a 
``reference" ($P_0$) and ``difference" ($P_1$) profile, respectively, and are related 
to an observed profile within this timespan using the relation $P = c_0P_0 + c_1P_1$. 
The coefficients $c_0$, $c_1$ were estimated using a cross-correlation algorithm 
between the observed profiles and principal components in the frequency domain, 
and the shape $F$ of each profile was then estimated by calculating the ratio 
$F = c_1/c_0$ in order to negate epoch-dependent scintillation effects.

The shape $F$ of a profile recorded at time $t$ and eccentric anomaly 
$u$ can also be determined using the relation 

\begin{equation}
  F_{\textrm{mod}}  = \frac{dF}{dt}t+\delta_\textrm{A}F(u) + \epsilon
  \label{eq:F}
\end{equation}

\noindent where $\epsilon$ is an intercept parameter and $dF/dt\textrm{ and }
\delta_\textrm{A}F$ are functions of the pulsar's precession rate 
$\Omega_1^{\textrm{spin}}$ and the angle $\eta$ between the line of 
nodes and the projection of the spin axis on the plane of the sky: 

\begin{eqnarray}
  \label{eq:dFdt}
  \frac{dF}{dt} &=& F'\Omega_1^{\textrm{spin}}\cos\eta\sin i \\
  \delta_\textrm{A}F &=& F'\frac{\beta_1}{\sin i}[-\cos\eta S(u)+\cos i\sin\eta C(u)]
  \label{eq:Faber}
\end{eqnarray}

\noindent The parameter $F' = dF/d\zeta$ characterizes the unknown beam 
structure as a function of the auxilary ``viewing" angle $\zeta$, $\beta_1 = 2\pi 
x/(P_b\sqrt{1-e^2})$ is the mean orbital velocity of the pulsar, and $C(u) = 
\cos[\omega+A_e(u)]+e\cos\omega$ and $S(u) = \sin[\omega+A_e(u)]+
e\sin\omega$ are time-dependent orbital terms that depend on the true anomaly 
$A_e(u)$. All binary parameters in Equations \ref{eq:dFdt} and \ref{eq:Faber} were 
determined through pulsar-timing techniques described in Section \ref{sec:timing}.

We fitted Equation \ref{eq:F} to our 430-MHz data using an MCMC implementation 
with a Metropolis algorithm in order to obtain posterior distributions of $\Omega_{1}
^{\textrm{spin}}$, $\eta$, $F'$, and $\epsilon$ from uniform priors. {\bfref We assumed that 
the joint posterior probability of the model $J (\Omega_1^{\textrm{spin}}, \eta, F', \epsilon | F, t, u)$ 
is a normal distribution in the $\chi^2$ goodness-of-fit statistic for the profile-shape 
model, 

\begin{equation}
  J \propto \textrm{ exp}\bigg[- 
   \frac{1}{2}\sum_i \bigg(\frac{F_i - F_{\textrm{mod}}(t_i,u_i)}{\sigma_i}\bigg)^2\bigg]
\end{equation}

\noindent where $\sigma_i$ is the uncertainty in $F_i$ determined from the cross 
correlation between the $i$th profile and the two principal components.} The results 
of this fitting procedure are summarized in Table \ref{tab:profdata} and discussed in 
Section \ref{sec:precres} below. 

\begin{figure}[h]
  \begin{center}
    \includegraphics[scale=0.43]{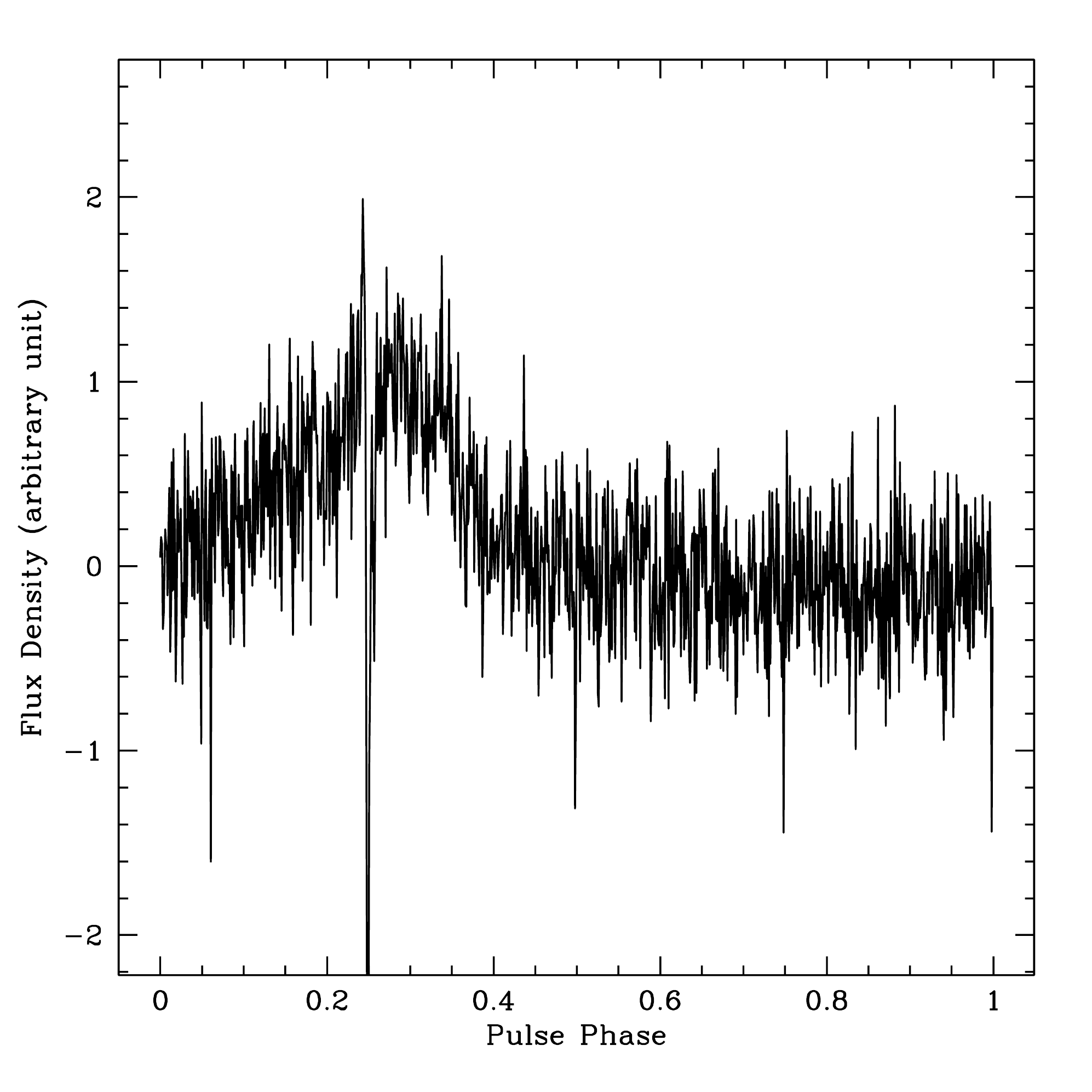}
    \caption{Difference between cumulative 2005 campaign profiles for the Mark IV and ASP 
    backends.}
    \label{fig:diff05}
  \end{center}
\end{figure}

As a second analysis, we fitted an RVM to available polarization position-angle 
data on each full-sum campaign profile. With the assumption of a dipolar magnetic 
geometry,  an RVM fit yields the angle between the pulsar's spin and magnetic axes 
($\alpha$), as well as the minimum-impact angle between the magnetic pole and line 
of sight ($\beta$). While no evolution is expected in $\alpha$, spin precession will cause 
$\beta$ to evolve with time such that $d\beta/dt = \Omega_{1}^{\textrm{spin}}\cos\eta\sin i$ 
\citep{dt92}. 

\noindent The combination of MCMC and RVM analyses therefore yields a test on 
observed profile evolution due to relativistic spin precession from two independent 
measurements. 

The differences in data quality between Mark IV and ASP profiles can be seen as slight 
differences in the profile shape across pulse phase, as shown in Figure \ref{fig:diff05}. This 
introduced slight discrepancies in the PCA results when performing a 
full analysis using all available data, which subsequently affected the derived profile 
shapes and MCMC results. Two separate studies between backends were not possible 
as the ASP era consisted of fewer profiles and a smaller timespan, with the 2008 campaign 
being excluded from the MCMC analysis due to having many low signal-to-noise profiles. We 
therefore decided to perform a PCA on all Mark IV profiles only, and then use the derived 
principal components to estimate the shapes for all high signal-to-noise Mark IV and ASP profiles. 
This approach does not account for observed scintillation or profile evolution across 
frequency in ASP data, and therefore only used ASP data collected with the two 
innermost frequency channels centered on 430 MHz for both analyses in order to 
minimize such effects.

\section{Discussion}
\label{sec:discuss}

\subsection{Pulse Jitter in PSR B1534+12}
\label{sec:jitter}

The measurability of $\ddot{\nu}$ and $\dddot{\nu}$ in PSR B1534+12 strongly 
suggests a significant amount of timing noise across our data set. This 
polynomial whitening in our best-fit model removes most of the long-term timing 
noise, which is usually attributed to rotational instabilities and variations in 
magnetospheric torque. However, TOA residuals generally exhibit scatter on shorter, 
pulse-period timescales in excess of standard measurement uncertainties. This 
residual ``jitter" is manifested from slight changes in the shape, amplitude and 
pulse phase of recorded profiles between successive pulses, and is likely due to 
variable activity within the pulsar magnetosphere. Recent studies of pulse jitter 
suggest that timing precision can be improved when averaging consecutive TOA  
residuals together \citep{sc12}.

\begin{figure}[h]
  \begin{center}
    \includegraphics[scale=0.43]{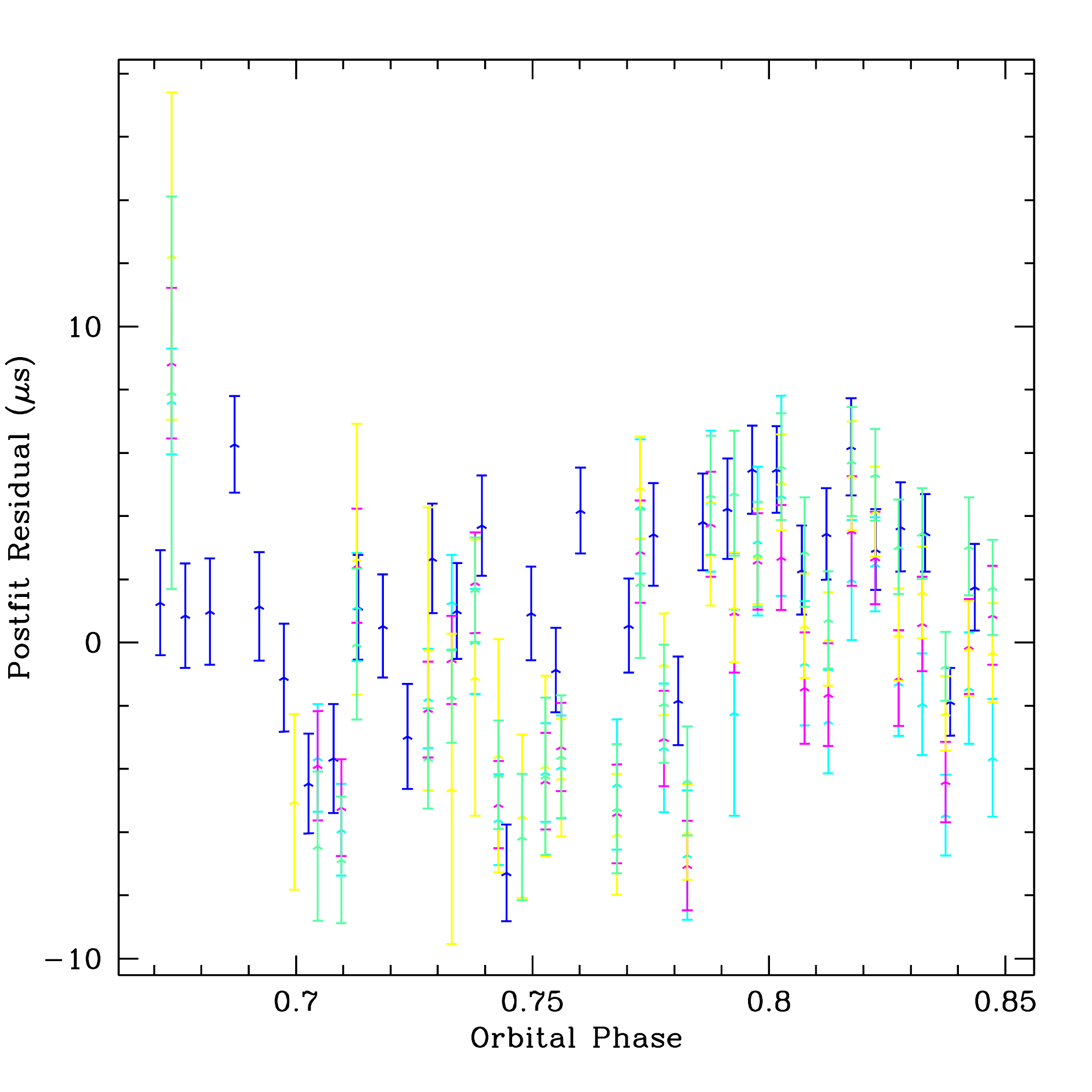}
    \caption{Pulse jitter in PSR B1534+12. The above figure shows global-fit 
    residuals of TOAs recorded on MJD 53545 as a function of orbital phase. The dark-blue 
    points were recorded with the Mark IV pulsar backend, and the remaining colors represent 
    the four channels of ASP 430-MHz data recorded simultaneously. }
    \label{fig:jitter}
  \end{center}
\end{figure}

We believe that such pulse jitter is evident in our timing analysis of PSR B1534+12. 
Figure \ref{fig:jitter} displays TOA residuals as a function of orbital phase from a 
global fit using unweighted TOA uncertainties recorded simultaneously on MJD 
53545. The dark-blue data points were recorded with the Mark IV backend, while 
additional points represent the four channelized ASP data sets. While residual 
variations are visibly uncorrelated within the shown timescale, there is a visible 
correlation between the two overall data sets despite significant differences in 
backend specifications. We therefore associate this backend-correlated scatter 
as pulse jitter due to irregular activity within the pulsar's magnetosphere.

Several implications arise from the observed pulse jitter. First, pulse jitter will 
become a significant source of timing error in future timing studies of PSR 
B1534+12. The recent installation of the PUPPI signal 
processor\footnote{\url{http://www.naic.edu/~astro/guide/node11.html}} is expected 
to produce high-precision residuals with scatter that strongly reflects time-dependent 
inhomogeneities in the magnetosphere. Second, a long-term solution to jitter with 
PSR B1534+12 cannot involve averaging a large number of consecutive TOAs. 
The main objective of strong-gravity tests with pulsars is to monitor time-dependent 
changes to orbital elements and quasi-static PK parameters, which requires full 
coverage of the orbit over long periods of time. We therefore use this jitter as a 
means to justify the TOA-uncertainty compensation for the global timing solution 
described in Section \ref{sec:obsred}. Lastly, further instrumental upgrades will 
only improve measurements made at 1400 MHz, where PSR B1534+12 is intrinsically 
weaker and signal-to-noise is currently limited. However, the overall timing solution 
is still expected to improve with additional observations over time.

\begin{deluxetable*}{lcc}
  \tabletypesize{\scriptsize}
  \tablecaption{Orbital Elements for PSR B1534+12}
  \tablewidth{0in}
  \tablehead{\colhead{Parameter} & \colhead{DD Model} & \colhead{DDGR Model}}
  \startdata
  Projected semimajor axis, \(x\) (s) \dotfill & 3.7294636(6) & 3.72946417(13) \\ 
  Eccentricity, \(e\) \dotfill & 0.27367752(7) & 0.27367740(4) \\
  Epoch of periastron, \(T_0\) (MJD) \dotfill & 52076.827113263(11) & 52076.827113271(9) \\
  Orbital Period, \(P_b\) (days) \dotfill & 0.420737298879(2) & 0.420737298881(2) \\ 
  Argument of periastron, \(\omega\) (deg) \dotfill & 283.306012(12) & 283.306029(10) \\
  & & \\
  Rate of periastron advance, \(\dot{\omega}\) (deg yr\(^{-1}\)) \dotfill & 1.7557950(19) & 
    1.755795\tablenotemark{a}  \\
  Time-averaged gravitational redshift, \(\gamma\) (ms) \dotfill & 2.0708(5) & 
    2.0701\tablenotemark{a} \\
  Orbital decay, \((\dot{P}_b)^{\textrm{obs}}\) (\(10^{-12}\)) \dotfill & -0.1366(3) & 
    -0.19244\tablenotemark{a} \\ 
  Shape of Shapiro delay, \(s = \textrm{sin}i\) \dotfill & 0.9772(16) & 0.97496\tablenotemark{a} \\
  Range of Shapiro delay, \(r = T_{\odot}m_2\) (\(\mu\)s) \dotfill & 6.6(2) & 
    6.627\tablenotemark{a}  \\
  & & \\
  Companion mass, \(m_2\textrm{ }(M_{\odot})\) \dotfill & 1.35(5) & 1.3455(2) \\ 
  Pulsar mass, \(m_1\textrm{ }(M_{\odot})\) \dotfill & $\cdots$ & 1.3330(2)\tablenotemark{a} \\
  Total mass, \(M = m_1+m_2 \textrm{ }(M_{\odot})\) \dotfill & $\cdots$ &  2.678463(4) \\
  Excess \(\dot{P}_b\textrm{ }(10^{-12})\) \dotfill & $\cdots$ & 0.0559(3) \\ 
  \enddata
  \label{tab:binarypar}
  \tablecomments{Values in parentheses denote the uncertainty in the preceding 
digit(s).}
  \tablenotetext{a}{derived quantity}
\end{deluxetable*}

\begin{figure}[h]
  \begin{center}
    \includegraphics[scale=0.43]{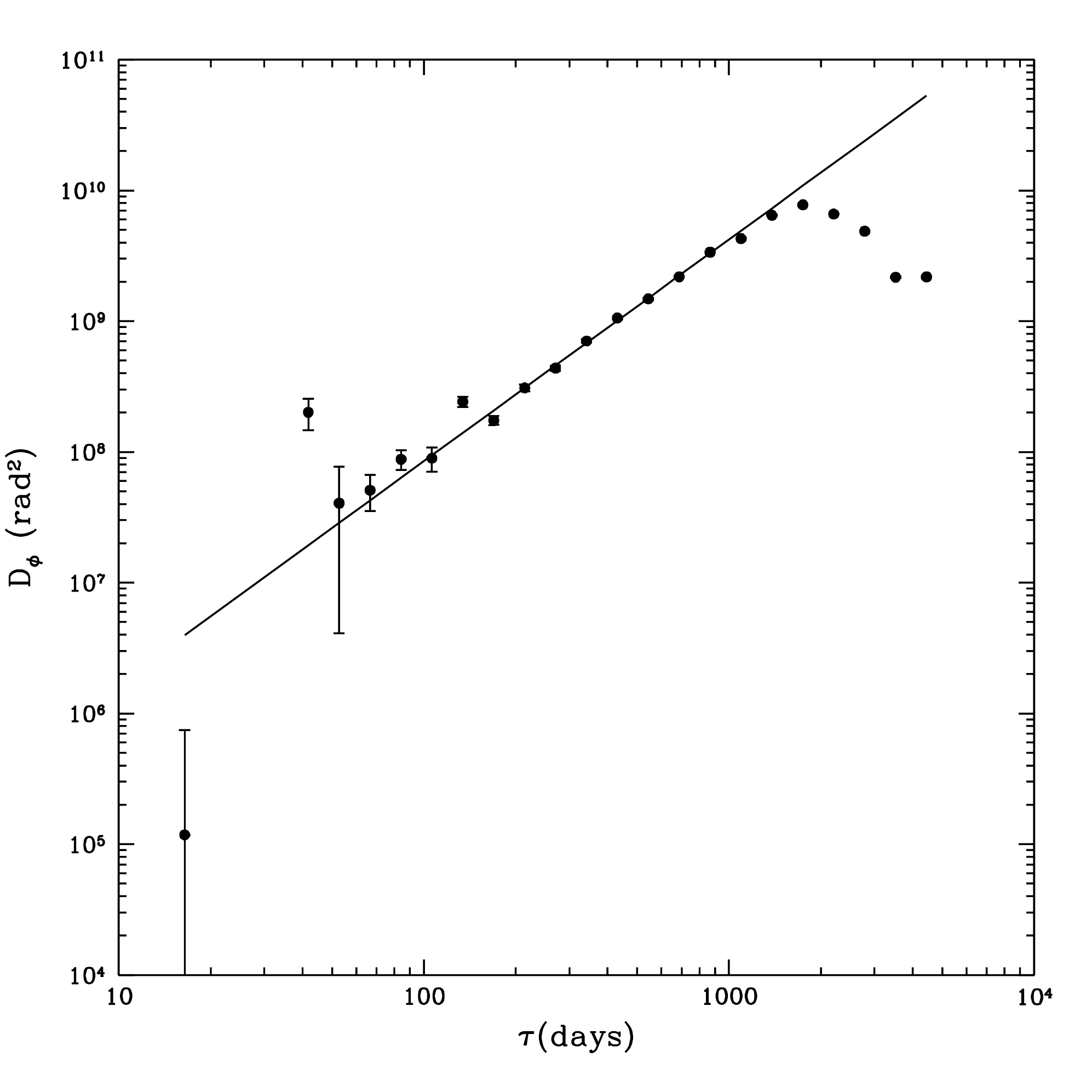}
    \caption{Phase structure function $D_{\phi}$ as a function of time lag $\tau$. The solid 
    line is a best-fit model of Equation \ref{eq:dmstructpred} for data with lags between   
    70 and 900 days.}
    \label{fig:dmstruct}
  \end{center}
\end{figure}

\subsection{Long-term Variations in Dispersion Measure}
\label{sec:vardm} 

Figure \ref{fig:dm} illustrates an irregular evolution in DM over time for PSR 
B1534+12. Five large bins, each with a fitted gradient, are used in our timing model 
to fully describe the observed changes across different timespans. Recent studies have 
shown that several objects also exhibit nonlinear evolution in DM along different 
directions and distances \citep[e.g.][]{kcs+13}. As with these studies, we believe that the 
dominant source of such evolution is the inhomogeneity of the interstellar medium, which 
is traced by the pulsar's signal as the line of sight sweeps through different regions due to 
a significant relative motion. 

{\bfref These long-term DM measurements are useful for a statistical analysis of 
turbulence within the interstellar medium \citep*[e.g.][]{ktr94}, which usually 
assumes that the power spectrum of spatial variations in electron density is 
a power law within a range of length scales \citep{ric90},

\begin{equation}
P(q) \propto q^{-\beta}, \hspace{20pt} q_o < q < q_i
\label{eq:pq}
\end{equation}

\noindent where $q = 2\pi/l$ is a spatial frequency and $l$ is a scattering length. 
The frequency range in Equation \ref{eq:pq} corresponds to a range between an ``inner" 
($l_i$) and ``outer" ($l_o$) length scale where the power-law form is valid. The observed 
spatial fluctuations due to a relative transverse velocity $v$ are related to a time lag $\tau$ 
by $l = v\tau$. The power spectrum $P(q)$ can therefore be estimated by computing a 
pulse-phase structure function $D_{\phi}(\tau) = \langle[\phi(t+\tau)-\phi(t)]^2\rangle$, 
where the angle brackets represent an ensemble average over observing epoch $t$. 
The pulse phase $\phi$ is linearly related to DM, which therefore relates $D_{\phi}(\tau)$ 
to a DM structure function $D_{\textrm{DM}}(\tau) = \langle[\textrm{DM}(t+\tau)-
\textrm{DM}(t)]^2\rangle$,

\begin{equation}
  D_{\phi}(\tau) = \bigg(\frac{2\pi C}{f}\bigg)^2D_{\textrm{DM}}(\tau)
\end{equation}
 
\noindent where $C = 4.148\times10^3 \textrm{ MHz}^2\textrm{ pc}^{-1}\textrm{ cm}^3$ 
s, and $f$ is the observing frequency in MHz. Moreover, $D_{\phi}(\tau)$ 
is a power law in $\tau$ within the inner length scales defined 
in Equation \ref{eq:pq}, which finally requires that 

\begin{equation}
  D_{\phi}(\tau) = \bigg(\frac{\tau}{\tau_0}\bigg)^{\beta-2}
  \label{eq:dmstructpred}
\end{equation}

\noindent where $\tau_0$ is a logarithmic intercept. Scintillation theory requires that 
$\tau_0 = \tau_\textrm{d}$, where $\tau_\textrm{d}$ is the diffractive timescale, if 
the inner length-scale $l_i \leq v\tau_{\textrm{d}}$.

We computed values of $D_{\phi}(\tau)$ at $f$ = 430 MHz using the Mark IV and 
ASP small-bin measurements of DM shown in Figure \ref{fig:dm}. The Mark III DM 
point was measured using all Mark III TOAs collected over several years, which were 
generated with a different standard profile than the one used for the Mark IV and ASP 
data; we therefore chose to ignore this measurement in order to avoid incorporating bias 
in the structure function. Uncertainties in $D_{\phi}(\tau)$ were determined by propagating 
errors from our DM($t$) measurements. Our estimate of $D_{\phi}(\tau)$ is  shown in Figure 
\ref{fig:dmstruct}, and illustrates a power-law evolution between time lags of roughly 70 and 
900 days. We fitted Equation \ref{eq:dmstructpred} to this segment of data, and found that 

\begin{align}
  \beta &= 3.70 \pm 0.04 \nonumber \\
  \tau_0 &= 3.0 \pm 0.8 \textrm{ minutes} 
  \label{eq:n8} 
\end{align}

\noindent which is shown as a solid black in in Figure \ref{fig:dmstruct}. 

The measured spectral index $\beta$ is consistent with the value for a ``Kolmogorov" 
medium, $\beta_{\rm Kol}$ = 11/3. Furthermore, $\beta$ and $\tau_0$ in Equation \ref{eq:n8} 
are consistent with the structure-function estimates reported by \citet{sw12}. Our estimate of 
$\tau_0$ is also consistent with the value of $\tau_{\textrm{d}}$ measured from the 
autocorrelation function of a dynamic spectrum of PSR B1534+12 \citep{bplw02}. 

At large timescales, the structure function departs from the fitted model at a lag $\tau_o 
\approx 900$ days, which suggests that 

\begin{equation}
  l_o \approx 52 \bigg(\frac{v}{100\textrm{ km/s}}\bigg)\textrm{ AU} 
  \label{eq:l_o}
\end{equation}

\noindent \citet{bplw02} derived an interstellar scintillation (ISS) velocity of 192 km/s. 
They noted in their study that ISS velocities of pulsars are dominated by the systemic 
transverse component, which means that $v \approx 192$ km/s for PSR B1534+12, 
and $l_o \sim 100$ AU $\sim 10^{15}$ cm from Equation \ref{eq:l_o}. This estimate is 
consistent with the upper limit of $l_o$ observed for several pulsars by \citet{pw91}. 
By contrast, there is no evidence for a significant inner scale from our data set, since 
bins with mean values less than 70 days contain only one or two pairs of DM$(t)$ and 
were therefore ignored from the analysis. We did not apply any correction for the 
solar-wind contribution of our DM($t$) measurements, due to a covariance between 
the TEMPO solar-wind DM model and a fitted timing parameter that is discussed at 
the end of Section \ref{sec:dist}.}

\subsection{High-precision Distance to PSR B1534+12}
\label{sec:dist}

Relative acceleration between the observatory and pulsar systems in the 
Galactic potential causes significant Doppler-factor biases in PK parameters. 
\citet{sac+98,sttw02} noted such behavior in earlier data sets of PSR B1534+12 
with the observed orbital decay and applied a distance-dependent kinematic 
correction derived by \citet{dt92} in order to include it as a consistent, radiative 
test of general relativity (see Section \ref{sec:tests} below). However, the corrected 
value yielded a large uncertainty due to the imprecise distance to the pulsar 
derived from DM measurements using the \citet{tc93} model of free electrons in 
the Galaxy. Previous studies of PSR B1534+12 therefore solved the inverse 
problem suggested by \citet{bb96}, where general relativity is assumed to be correct; 
the distance is then derived using the measured ``excess" orbital decay, a model 
of the Galactic acceleration \citep{kg89}, and the expression for kinematic correction 
derived by \citet{dt92}. Using this procedure, \citet{sttw02} were able to derive a 
distance with a relative uncertainty of 4.9\% when doubling their TEMPO uncertainties.

We used the same approach in this study to update the derived distance with a 
substantially longer timespan of Arecibo data. We also corrected the expression for 
the kinematic bias presented in \citet{sac+98,sttw02} for missing factors of the cosine 
of the pulsar's galactic latitude; the correct equation is given by \citet[Equation 5]{nt95}. Our 
derived distance to PSR B1534+12, using our timing results,  the corrected kinematic 
equation and updated Galactic parameters from \citet{rmb+14}, is 

\begin{figure}[h]
  \begin{center}
    \includegraphics[scale=0.43]{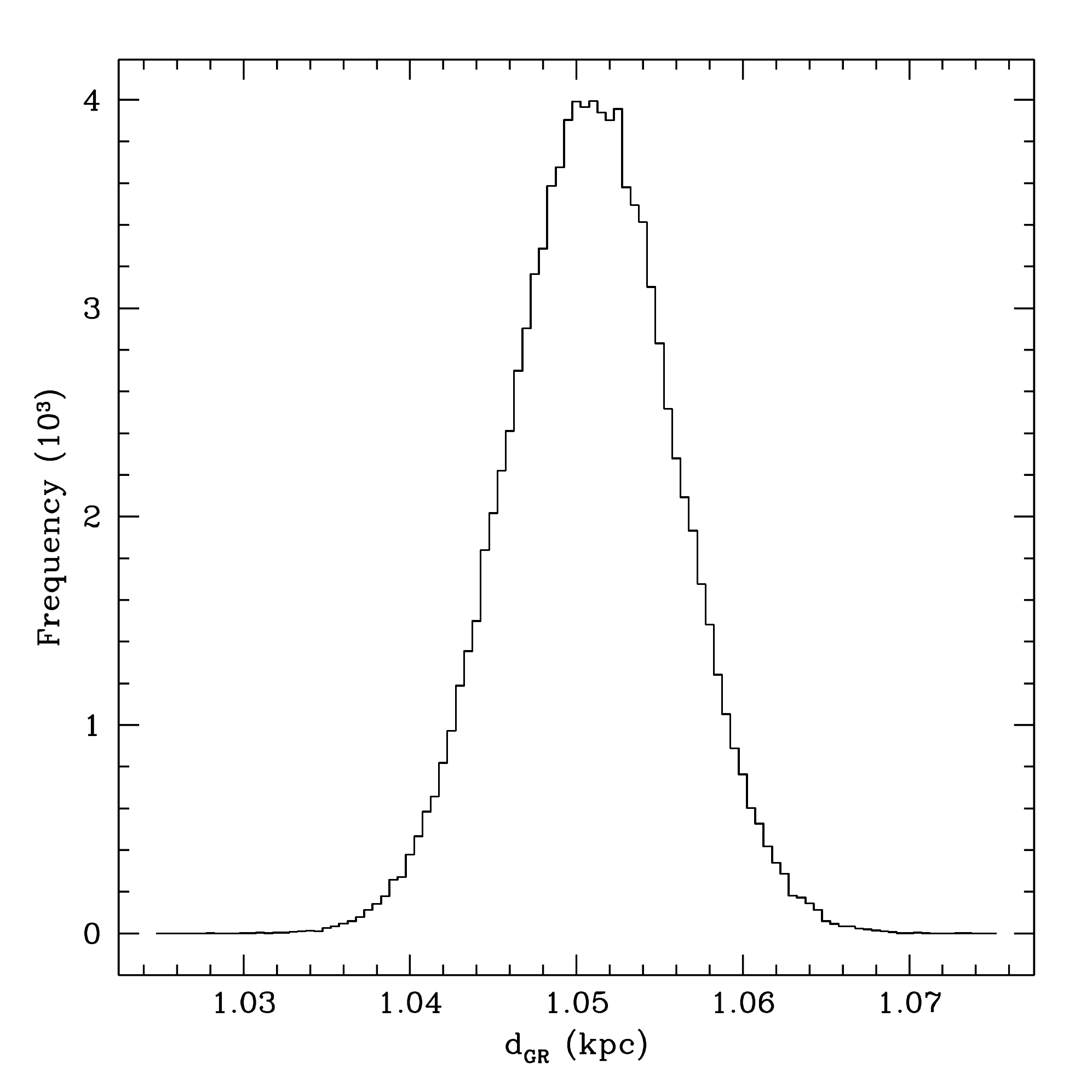}
    \caption{Distribution of $d_{\textrm{GR}}$ obtained by using a Monte-Carlo method 
    described in Section \ref{sec:dist}.}
    \label{fig:histdist}
  \end{center}
\end{figure}

{\bfref 
\begin{equation}
  d_{\textrm{GR}} = 1.051 \pm 0.005 \textrm{ kpc}
  \label{eq:distance}
\end{equation}

\noindent where the value and its uncertainty (68\% confidence level) were 
estimated using a Monte-Carlo method: all uncertain parameters were randomly 
sampled from a normal distribution with mean and standard deviation equal to their 
fitted values and uncertainties, respectively, which were then used to derive a value 
of $d_{\textrm{GR}}$ by applying Newton-Raphson's method \citep[e.g.][]{pftv86}. 
This process was repeated $10^5$ times, and resulted in a distribution of 
$d_{\textrm{GR}}$ that is shown in Figure \ref{fig:histdist}.} This new distance is 
consistent with the previous estimate of $1.02 \pm 0.05$ kpc made by \citet{sttw02}. 
The relative uncertainty of this result (0.48\%) is slightly 
lower than that of the derived distance to PSR J0437-4715 estimated by \citet{vbv+08}. We 
attribute this improvement in precision to the updated $(\dot{P}_b)^{\textrm{obs}}$ 
listed in Table \ref{tab:binarypar}. The uncertainty in $d_{\textrm{GR}}$ is 
dominated by uncertainties in Galactic acceleration and rotation parameters 
used to derive the estimate.

Despite its high level of precision, the derived distance presented in 
Equation \ref{eq:distance} is a model-dependent quantity. An ideal measure 
of distance can be obtained from the geometric, model-independent parallax 
through low-frequency interferometry. The recent inclusion of PSR B1534+12 
into an extension of the PSR$\pi$ interferometry 
program\footnote{\url{https://safe.nrao.edu/vlba/psrpi}} 
will likely provide such an estimate in the next 2-3 years. Another independent 
distance measure can be derived from a timing parallax estimated in the global-fit 
timing solution. However, our measured timing parallax was found to be significantly 
covariant with an input DM parameter associated with free electrons from the solar 
wind. Such a covariance is unexpected since the solar-wind component of DM is 
strongest for pulsars close to the ecliptic plane, while PSR B1534+12 is $\sim 30^{\circ}$ 
above the plane. Since the expected solar contribution is much smaller than the 
scatter of the 80-day DM bins in Figure \ref{fig:dm}, we chose to set the solar DM 
component to zero for the global timing fit while acknowledging that the timing parallax is 
unreliable as a fitted parameter.

 \begin{figure}[h]
  \begin{center}
    \includegraphics[scale=0.43]{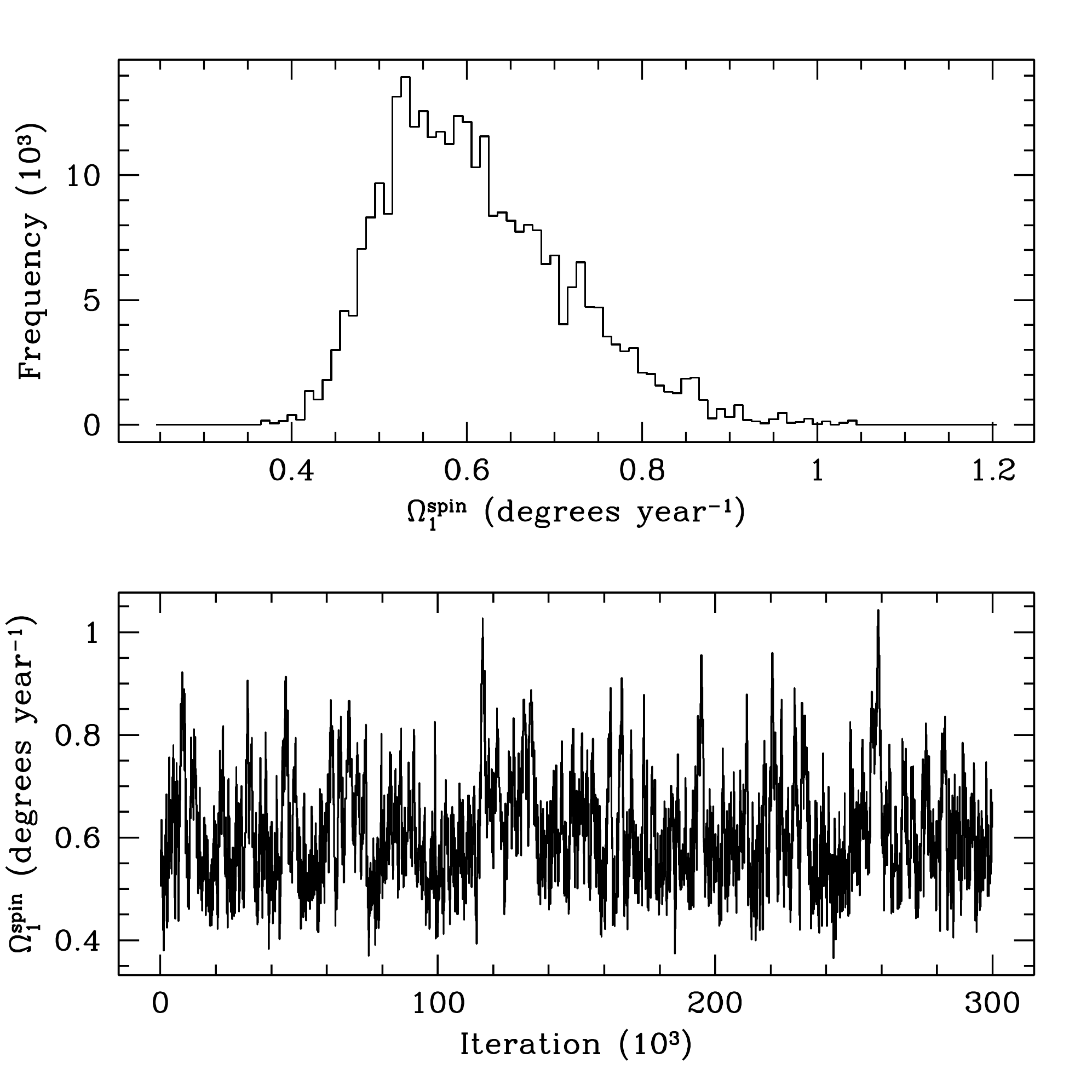}
    \caption{\emph{Top}: MCMC posterior distribution of $\Omega_1^{\textrm{spin}}$ 
      obtained from the profile-shape analysis of Mark IV and ASP data discussed in 
      Section \ref{sec:precres}. \emph{Bottom}: Markov chain for $\Omega_1^{\textrm{spin}}$
      determined from the MCMC algorithm.}
    \label{fig:mcmcpos}
  \end{center}
\end{figure}

\subsection{Precession and Geometry}
\label{sec:precres}

Results from the MCMC fit on several data sets can be found in Table \ref{tab:profdata}, 
and the posterior distribution for $\Omega_1^{\textrm{spin}}$ derived from our Mark IV 
and ASP data sets is shown in Figure \ref{fig:mcmcpos}. {\bfref We generated $3\times10^5$ 
samples for each application of the algorithm, after burning the first 5000 samples 
in order to remove non-convergent iterations.} We provided the original results 
obtained by STA04, as well as a reproduced set of results from the STA04 data set using 
the MCMC algorithm, for comparison with our extended Mark IV and ASP profiles. We 
assumed that values of $F'$ must be negative while using the MCMC algorithm, since 
the simultaneous-linear-fit technique used and described by STA04, which avoids any 
consideration of $F'$, estimates that $\cos\eta < 0$. These results agree well with predictions 
from general relativity, where $\Omega_1^{\textrm{spin}} = 0.51^{\circ}$/yr using the derived 
masses in Table \ref{tab:binarypar}, and previous measurements made by STA04. General 
improvements in precision come from the new fitting procedure, which permitted direct 
sampling of the precession rate and other free parameters, as well as the addition of the 
ASP 2005 campaign and several strong bi-monthly observations. 

\begin{deluxetable}{lcccc}
  \tabletypesize{\scriptsize}
  \tablewidth{0pt}
  \tablecaption{Profile-evolution MCMC Parameters}  
  \tablehead{\colhead{Parameter} & \colhead{STA04\tablenotemark{a}} & 
    \colhead{STA04} & \colhead{Mark IV} & 
    \colhead{All}}
  \startdata
  $\Omega_1^{\textrm{spin}}$ $(^{\circ}/\textrm{yr})$ \dotfill & $0.44^{+0.48}_{-0.16}$ & $0.51^{+0.10}_{-0.08}$ & $0.48^{+0.09}_{-0.07}$ & $0.59^{+0.12}_{-0.08}$ \\ [5pt]
  $\eta$ $(^{\circ})$ \dotfill & $\pm103^{+10}_{-10}$ & $\pm99^{+2}_{-2}$ & $\pm118^{+10}_{-15}$ & $\pm139^{+16}_{-25}$ \\ [5pt]
  $F'$ \dotfill & n/a & $-5.9^{+0.9}_{-1.0}$ & $-2.2^{+0.6}_{-0.7}$ & $-1.3^{+0.3}_{-0.5}$ \\ [5pt]
  $\epsilon$ ($10^{-3}$) \dotfill & $-1.5^{+0.3}_{-0.3}$ & $-1.90^{+0.08}_{-0.09}$  & $-1.21^{+0.08}_{-0.08}$ & $6.67^{+0.07}_{-0.07}$ \\ [5pt]
  \enddata
  \label{tab:profdata}
  \tablenotetext{a}{Original, non-MCMC results from STA04.}
  \tablecomments{Uncertainties reflect 68\% confidence intervals of posterior distributions.}
 \end{deluxetable}
 
\begin{figure}[h]
  \begin{center}
    \includegraphics[scale=0.43]{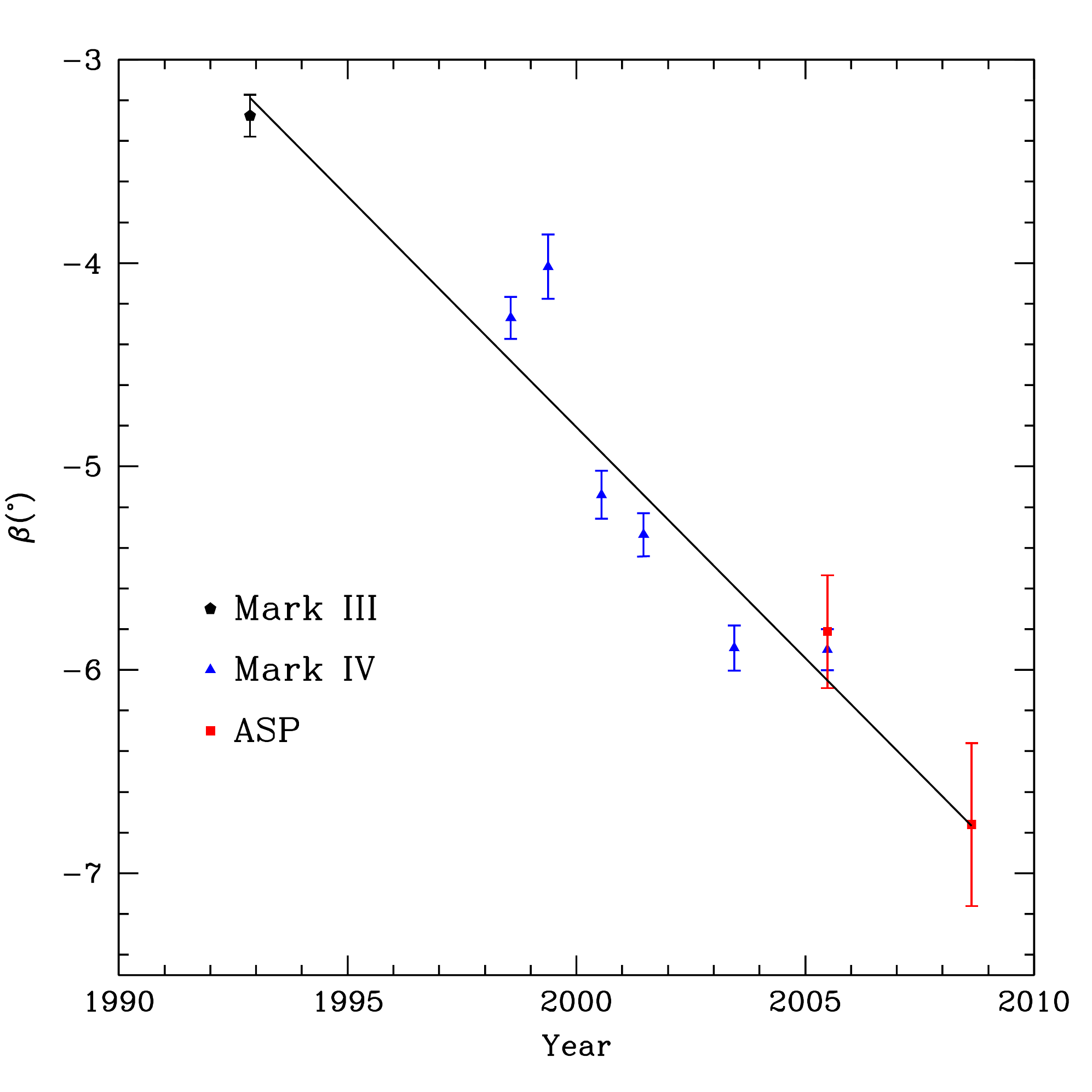}
    \caption{Impact angle $\beta$ between the magnetic axis and line of sight as a 
    function of time. The black line is a best-fit slope of -0.23 $\pm$ 0.02 $^{\circ}$/yr.}
    \label{fig:betavtime}
  \end{center}
\end{figure}

The RVM analysis yielded values of $\alpha$ and $\beta$ at different times using the 
Mark III (reticon), Mark IV and ASP campaign profiles. The values of $\beta$ measured 
for each campaign are shown in Figure \ref{fig:betavtime}. Measurements of $\alpha = 
103.5(3)^{\circ}$ are consistent with no evolution in time, while the values of $\beta$ are 
found to change significantly, where $d\beta/dt$ = -0.23 $\pm$ 0.02 $^{\circ}$/yr. This 
is consistent with the STA04 result of -0.21 $\pm$ 0.03 $^{\circ}$/yr. The 
assumption that general relativity is correct requires that $d\beta/dt = \Omega_1^{\textrm{spin}} 
\sin i\cos\eta$, and therefore yields $\eta = \pm 117 \pm 3^{\circ}$ (68\% confidence), 
which agrees with the value determined from the MCMC analysis described above. 
With these values, the misalignment angle $\delta$ between the spin and orbital angular 
momentum axes can be derived through  spherical trigonometry by 
$\cos\delta = -\sin i\sin\lambda\sin\eta+\cos\lambda\cos i$. 
The sign ambiguity in $\eta$ and $i$, as well as the requirement that $\cos i\tan\eta > 0$ 
pointed out by STA04, gives an expected value of $\delta = 27.0 \pm 3.0^{\circ}$ 
or $\delta = 153.0 \pm 3.0^{\circ}$. Physical arguments based on alignment of angular 
momenta prior to the second supernova suggest that the smaller angle is correct \citep{bai88}, 
and therefore requires that $\eta = -117 \pm 3^{\circ}$ and $i = 77.7 \pm 0.9^{\circ}$.

The consistency between the MCMC and RVM analyses serves as an improved, 
independent check of precession within this relativistic binary system. These results also 
confirm the geometric picture of this pulsar-binary system derived in STA04.

\citet{kws03} pointed out that relativistic spin precession can eventually imprint 
a timing signature as a second-derivative in spin frequency. However, they predicted 
that it would take an additional 25 years of timing observations in order to measure the 
predicted signature with reasonable accuracy. Moreover, our best-fit timing model of 
PSR B1534+12 data indicates that the measured $\ddot{\nu}$ and $\dddot{\nu}$ 
are dominated by timing noise.  

\begin{figure}[h]
  \begin{center}
    \includegraphics[scale=0.43]{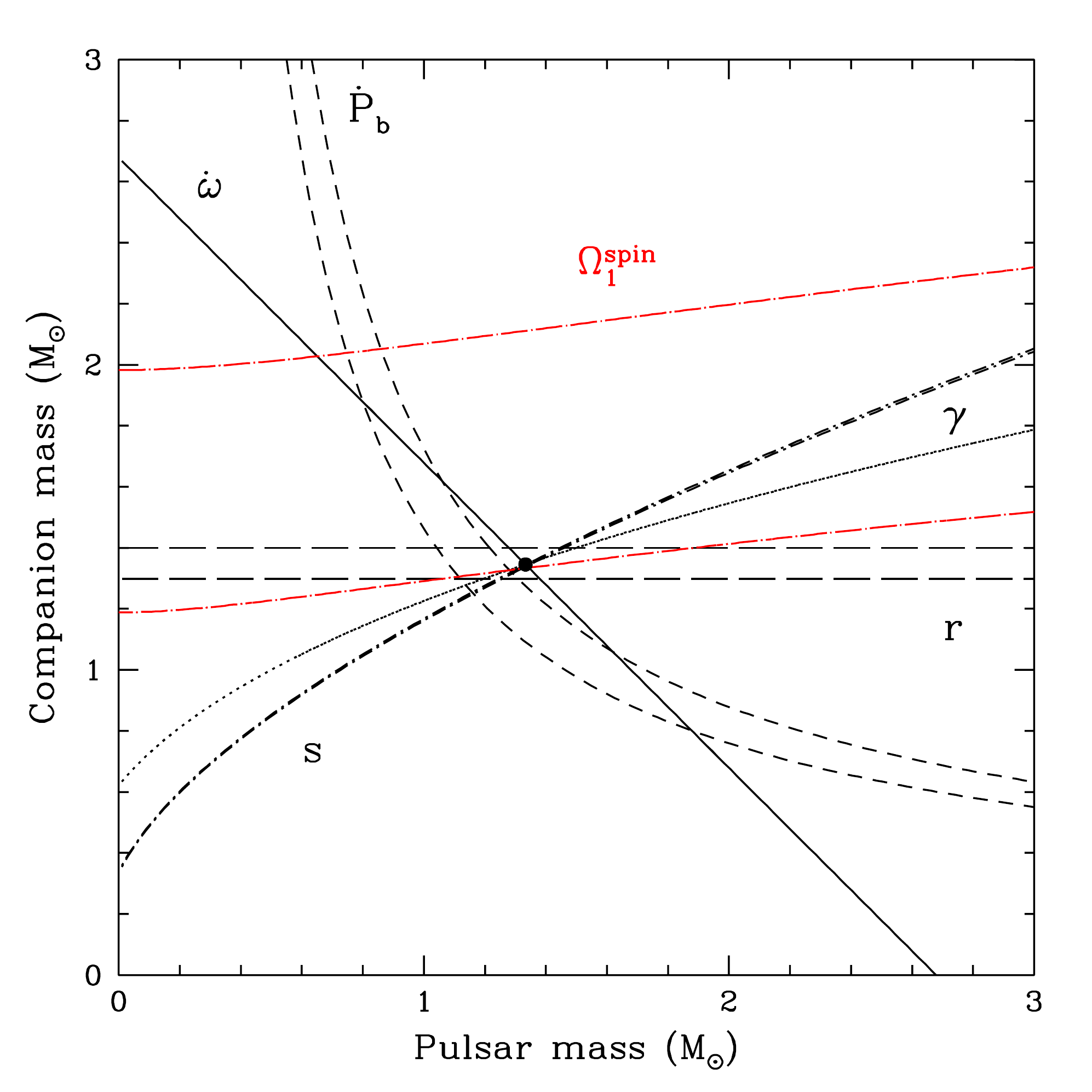}
    \caption{Mass-mass plot for PSR B1534+12. Each set of black curves represents the 
      68\%-confidence region delimited by the labeled PK (timing) parameter. The dot-dashed 
      red curves represent the 68\% confidence region determined from 
      the spin-precession rate. }
    \label{fig:m1m2}
  \end{center}
\end{figure}

\subsection{Tests of General Relativity} 
\label{sec:tests}

In general relativity, each PK parameter is expressed as a function of at least 
one of the two binary-component masses; one can therefore define a ``mass-mass" 
space where each PK parameter corresponds to a curve in this plane.  \citet{sac+98} 
presented the first mass-mass plot that incorporated up to 5 PK curves,  as well as 
the first ``non-mixed" test of quasi-static parameters using the $\dot{\omega}-\gamma-s$ 
combination. 

Figure \ref{fig:m1m2} presents our evaluation of PK parameters and tests of general 
relativity using PSR B1534+12. Each curve corresponds to a PK value measured either 
with the best-fit DD timing model or profile-shape model, while the filled circle represents the 
best-fit DDGR solution of the pulsar and companion masses of $m_1 = 1.3330\textrm{ }M_{\odot}$, 
$m_2 = 1.3455\textrm{ }M_{\odot}$, respectively. The $\dot{P}_b$ curve was corrected 
for kinematic bias assuming a pulsar distance of $d = 0.7 \pm 0.2$ kpc, which was estimated 
using the electron number-density model developed by \citet{tc93}, as discussed in Section 
\ref{sec:dist}. The discrepancy between this distance and the expected distance (Equation 
\ref{eq:distance}) prevents the curve from intersecting the other PK curves, while its large 
uncertainty dominates the corrected orbital-decay error estimate. The 
$\Omega_1^{\textrm{spin}}$ curves intersect the best-fit DDGR point well at the 68\% 
confidence level.

The Shapiro $r$ parameter remains a slightly weaker constraint than the 
PK timing parameters, but its relative uncertainty has improved by nearly a factor of 
two since the last measurement by \citet{sttw02}. The $\dot{\omega}-\gamma-s$ combination 
is the strongest test from this pulsar and confirms general relativity to within 0.17\% of its 
predictions. This test quality is slightly larger than the 0.05\% test from the double-pulsar 
system, which uses the mass ratio as determined by the projected semi-major axes of both 
pulsars \citep{ksm+06}. 

The high-precision DDGR masses of PSR B1534+12 and its companion are 
consistent with previous estimates. As noted in \citet{sttw02}, the significant difference 
between $m_1$ and $m_2$ presents a conundrum where the spun-up pulsar is 
actually less massive than its companion. This suggests that a period of ``mass inversion" 
 -- mass transfer from the pulsar's progenitor to its companion -- took place during the system's 
 evolution, though a more thorough understanding of mass-transfer processes and its 
 effects on stellar structure is needed. The effectiveness of mass estimates from Shapiro-delay 
 measurements will continue to provide a better view of the general pulsar-companion mass 
 population \citep{kkdt13}.\\

The Arecibo Observatory is operated by SRI International under a cooperative 
agreement with the National Science Foundation (AST-1100968), and in 
alliance with Ana G. M\'{e}ndez-Universidad Metropolitana, and the Universities 
Space Research Association. Pulsar research at UBC is supported by an NSERC 
Discovery Grant. We thank Z. Arzoumanian, F. Camilo, A. Lyne, 
D. Nice, J. H. Taylor, and A. Wolszczan for their earlier contributions to this 
project. We also thank P. Freire, I. Hoffman, A. Lommen, D. Lorimer, D. Nice, 
E. Splaver, and K. Xilouris for previous assistance with Mark-IV observations. 
We also thank R. Ferdman, M. Gonzalez, other NANOGrav observers for help 
with some of the ASP observations reported here. We thank W. Zhu for providing 
a TEMPO MCMC program. We are grateful for useful comments provided by an 
anonymous referee.

\bibliography{paperforarXiv}

\begin{thebibliography}{45}
\expandafter\ifx\csname natexlab\endcsname\relax\def\natexlab#1{#1}\fi

\bibitem[{{Antoniadis} {et~al.}(2013){Antoniadis}, {Freire}, {Wex}, {Tauris},
  {Lynch}, {van Kerkwijk}, {Kramer}, {Bassa}, {Dhillon}, {Driebe}, {Hessels},
  {Kaspi}, {Kondratiev}, {Langer}, {Marsh}, {McLaughlin}, {Pennucci}, {Ransom},
  {Stairs}, {van Leeuwen}, {Verbiest}, \& {Whelan}}]{afw+13}
{Antoniadis}, J., {Freire}, P.~C.~C., {Wex}, N., {Tauris}, T.~M., {Lynch},
  R.~S., {van Kerkwijk}, M.~H., {Kramer}, M., {Bassa}, C., {Dhillon}, V.~S.,
  {Driebe}, T., {Hessels}, J.~W.~T., {Kaspi}, V.~M., {Kondratiev}, V.~I.,
  {Langer}, N., {Marsh}, T.~R., {McLaughlin}, M.~A., {Pennucci}, T.~T.,
  {Ransom}, S.~M., {Stairs}, I.~H., {van Leeuwen}, J., {Verbiest}, J.~P.~W., \&
  {Whelan}, D.~G. 2013, Science, 340

\bibitem[{Arzoumanian(1995)}]{arz95}
Arzoumanian, Z. 1995, PhD thesis, Princeton University

\bibitem[{Bailes(1988)}]{bai88}
Bailes, M. 1988, A\&A, 202, 109

\bibitem[{Bell \& Bailes(1996)}]{bb96}
Bell, J.~F. \& Bailes, M. 1996, ApJ, 456, L33

\bibitem[{Bogdanov {et~al.}(2002)Bogdanov, Pruszunska, Lewandowski, \&
  Wolszczan}]{bplw02}
Bogdanov, S., Pruszunska, M., Lewandowski, W., \& Wolszczan, A. 2002, ApJ, 581,
  495

\bibitem[{{Breton} {et~al.}(2008){Breton}, {Kaspi}, {Kramer}, McLaughlin,
  Lyutikov, Ransom, Stairs, Ferdman, \& Camilo}]{bkk+08}
{Breton}, R.~P., {Kaspi}, V.~M., {Kramer}, M., McLaughlin, M.~A., Lyutikov, M.,
  Ransom, S.~R., Stairs, I.~H., Ferdman, R.~D., \& Camilo, F. 2008, Science

\bibitem[{{Burgay} {et~al.}(2003){Burgay}, {D'Amico}, {Possenti}, {Manchester},
  {Lyne}, {Joshi}, {McLaughlin}, {Kramer}, {Sarkissian}, {Camilo}, {Kalogera},
  {Kim}, \& {Lorimer}}]{bdp+03}
{Burgay}, M., {D'Amico}, N., {Possenti}, A., {Manchester}, R.~N., {Lyne},
  A.~G., {Joshi}, B.~C., {McLaughlin}, M.~A., {Kramer}, M., {Sarkissian},
  J.~M., {Camilo}, F., {Kalogera}, V., {Kim}, C., \& {Lorimer}, D.~R. 2003,
  Nature, 426, 531

\bibitem[{Damour \& Deruelle(1985)}]{dd85}
Damour, T. \& Deruelle, N. 1985, Ann. Inst. H. Poincar\'e (Physique
  Th\'eorique), 43, 107

\bibitem[{Damour \& Deruelle(1986)}]{dd86}
---. 1986, Ann. Inst. H. Poincar\'e (Physique Th\'eorique), 44, 263

\bibitem[{Damour \& Taylor(1992)}]{dt92}
Damour, T. \& Taylor, J.~H. 1992, Phys. Rev. D, 45, 1840

\bibitem[{{de Sitter}(1916)}]{ds16}
{de Sitter}, W. 1916, MNRAS, 77, 155

\bibitem[{Demorest(2007)}]{dem07}
Demorest, P.~B. 2007, PhD thesis, University of California, Berkeley

\bibitem[{Fonseca(2012)}]{fon12}
Fonseca, E. 2012, Master's thesis, The University of British Columbia,
  \url{https://circle.ubc.ca/handle/2429/43498}

\bibitem[{{Gregory}(2005)}]{gre05bayes}
{Gregory}, P.~C. 2005, {Bayesian Logical Data Analysis for the Physical
  Sciences: A Comparative Approach with `Mathematica' Support} (Cambridge
  University Press)

\bibitem[{{Hankins} \& {Rickett}(1975)}]{hr75}
{Hankins}, T.~H. \& {Rickett}, B.~J. 1975, in Methods in Computational Physics
  Volume 14 --- Radio Astronomy (New York: Academic Press), 55

\bibitem[{Hulse \& Taylor(1975)}]{ht75a}
Hulse, R.~A. \& Taylor, J.~H. 1975, ApJ, 195, L51

\bibitem[{Kaspi {et~al.}(1994)Kaspi, Taylor, \& Ryba}]{ktr94}
Kaspi, V.~M., Taylor, J.~H., \& Ryba, M. 1994, ApJ, 428, 713

\bibitem[{{Keith} {et~al.}(2013){Keith}, {Coles}, {Shannon}, {Hobbs},
  {Manchester}, {Bailes}, {Bhat}, {Burke-Spolaor}, {Champion}, {Chaudhary},
  {Hotan}, {Khoo}, {Kocz}, {Os{\l}owski}, {Ravi}, {Reynolds}, {Sarkissian},
  {van Straten}, \& {Yardley}}]{kcs+13}
{Keith}, M.~J., {Coles}, W., {Shannon}, R.~M., {Hobbs}, G.~B., {Manchester},
  R.~N., {Bailes}, M., {Bhat}, N.~D.~R., {Burke-Spolaor}, S., {Champion},
  D.~J., {Chaudhary}, A., {Hotan}, A.~W., {Khoo}, J., {Kocz}, J.,
  {Os{\l}owski}, S., {Ravi}, V., {Reynolds}, J.~E., {Sarkissian}, J., {van
  Straten}, W., \& {Yardley}, D.~R.~B. 2013, MNRAS, 429, 2161

\bibitem[{{Kiziltan} {et~al.}(2013){Kiziltan}, {Kottas}, {De Yoreo}, \&
  {Thorsett}}]{kkdt13}
{Kiziltan}, B., {Kottas}, A., {De Yoreo}, M., \& {Thorsett}, S.~E. 2013, ApJ,
  778, 66

\bibitem[{{Konacki} {et~al.}(2003){Konacki}, {Wolszczan}, \& {Stairs}}]{kws03}
{Konacki}, M., {Wolszczan}, A., \& {Stairs}, I.~H. 2003, ApJ, 589, 495

\bibitem[{{Kramer}(1998)}]{kra98}
{Kramer}, M. 1998, ApJ, 509, 856

\bibitem[{{Kramer} {et~al.}(2006){Kramer}, {Stairs}, {Manchester},
  {McLaughlin}, {Lyne}, {Ferdman}, {Burgay}, {Lorimer}, {Possenti}, {D'Amico},
  {Sarkissian}, {Hobbs}, {Reynolds}, {Freire}, \& {Camilo}}]{ksm+06}
{Kramer}, M., {Stairs}, I.~H., {Manchester}, R.~N., {McLaughlin}, M.~A.,
  {Lyne}, A.~G., {Ferdman}, R.~D., {Burgay}, M., {Lorimer}, D.~R., {Possenti},
  A., {D'Amico}, N., {Sarkissian}, J.~M., {Hobbs}, G.~B., {Reynolds}, J.~E.,
  {Freire}, P.~C.~C., \& {Camilo}, F. 2006, Science, 314, 97

\bibitem[{Kuijken \& Gilmore(1989)}]{kg89}
Kuijken, K. \& Gilmore, G. 1989, MNRAS, 239, 571

\bibitem[{Lorimer \& Kramer(2005)}]{lk05}
Lorimer, D.~R. \& Kramer, M. 2005, Handbook of Pulsar Astronomy (Cambridge
  University Press)

\bibitem[{{Manchester} {et~al.}(2010){Manchester}, {Kramer}, {Stairs},
  {Burgay}, {Camilo}, {Hobbs}, {Lorimer}, {Lyne}, {McLaughlin}, {McPhee},
  {Possenti}, {Reynolds}, \& {van Straten}}]{mks+10}
{Manchester}, R.~N., {Kramer}, M., {Stairs}, I.~H., {Burgay}, M., {Camilo}, F.,
  {Hobbs}, G.~B., {Lorimer}, D.~R., {Lyne}, A.~G., {McLaughlin}, M.~A.,
  {McPhee}, C.~A., {Possenti}, A., {Reynolds}, J.~E., \& {van Straten}, W.
  2010, ApJ, 710, 1694

\bibitem[{Nice \& Taylor(1995)}]{nt95}
Nice, D.~J. \& Taylor, J.~H. 1995, ApJ, 441, 429

\bibitem[{Phillips \& Wolszczan(1991)}]{pw91}
Phillips, J.~A. \& Wolszczan, A. 1991, ApJ, 382, L27

\bibitem[{Press {et~al.}(1986)Press, Flannery, Teukolsky, \&
  Vetterling}]{pftv86}
Press, W.~H., Flannery, B.~P., Teukolsky, S.~A., \& Vetterling, W.~T. 1986,
  Numerical Recipes: {T}he Art of Scientific Computing (Cambridge: Cambridge
  University Press)

\bibitem[{Radhakrishnan \& Cooke(1969)}]{rc69a}
Radhakrishnan, V. \& Cooke, D.~J. 1969, Astrophys. Lett., 3, 225

\bibitem[{{Reid} {et~al.}(2014){Reid}, {Menten}, {Brunthaler}, {Zheng}, {Dame},
  {Xu}, {Wu}, {Zhang}, {Sanna}, {Sato}, {Hachisuka}, {Choi}, {Immer},
  {Moscadelli}, {Rygl}, \& {Bartkiewicz}}]{rmb+14}
{Reid}, M.~J., {Menten}, K.~M., {Brunthaler}, A., {Zheng}, X.~W., {Dame},
  T.~M., {Xu}, Y., {Wu}, Y., {Zhang}, B., {Sanna}, A., {Sato}, M., {Hachisuka},
  K., {Choi}, Y.~K., {Immer}, K., {Moscadelli}, L., {Rygl}, K.~L.~J., \&
  {Bartkiewicz}, A. 2014, ArXiv e-prints

\bibitem[{Rickett(1990)}]{ric90}
Rickett, B.~J. 1990, Ann. Rev. Astr. Ap., 28, 561

\bibitem[{{Scheiner} \& {Wolszczan}(2012)}]{sw12}
{Scheiner}, B. \& {Wolszczan}, A. 2012, in American Astronomical Society
  Meeting Abstracts, Vol. 220, American Astronomical Society Meeting Abstracts
  \#220, 430.06

\bibitem[{{Shannon} \& {Cordes}(2012)}]{sc12}
{Shannon}, R.~M. \& {Cordes}, J.~M. 2012, apj, 761, 64

\bibitem[{Stairs {et~al.}(1998)Stairs, Arzoumanian, Camilo, Lyne, Nice, Taylor,
  Thorsett, \& Wolszczan}]{sac+98}
Stairs, I.~H., Arzoumanian, Z., Camilo, F., Lyne, A.~G., Nice, D.~J., Taylor,
  J.~H., Thorsett, S.~E., \& Wolszczan, A. 1998, ApJ, 505, 352

\bibitem[{Stairs {et~al.}(2000)Stairs, Splaver, Thorsett, Nice, \&
  Taylor}]{sst+00}
Stairs, I.~H., Splaver, E.~M., Thorsett, S.~E., Nice, D.~J., \& Taylor, J.~H.
  2000, MNRAS, 314, 459, (astro-ph/9912272)

\bibitem[{{Stairs} {et~al.}(2004){Stairs}, {Thorsett}, \&
  {Arzoumanian}}]{sta04}
{Stairs}, I.~H., {Thorsett}, S.~E., \& {Arzoumanian}, Z. 2004, Phys. Rev.
  Lett., 93, 141101

\bibitem[{Stairs {et~al.}(2002)Stairs, Thorsett, Taylor, \& Wolszczan}]{sttw02}
Stairs, I.~H., Thorsett, S.~E., Taylor, J.~H., \& Wolszczan, A. 2002, ApJ, 581,
  501

\bibitem[{Stinebring {et~al.}(1992)Stinebring, Kaspi, Nice, Ryba, Taylor,
  Thorsett, \& Hankins}]{skn+92}
Stinebring, D.~R., Kaspi, V.~M., Nice, D.~J., Ryba, M.~F., Taylor, J.~H.,
  Thorsett, S.~E., \& Hankins, T.~H. 1992, Rev. Sci. Instrum., 63, 3551

\bibitem[{Taylor(1992)}]{tay92}
Taylor, J.~H. 1992, Philos. Trans. Roy. Soc. London A, 341, 117

\bibitem[{Taylor \& Cordes(1993)}]{tc93}
Taylor, J.~H. \& Cordes, J.~M. 1993, ApJ, 411, 674

\bibitem[{Taylor {et~al.}(1992)Taylor, Wolszczan, Damour, \& Weisberg}]{twdw92}
Taylor, J.~H., Wolszczan, A., Damour, T., \& Weisberg, J.~M. 1992, Nature, 355,
  132

\bibitem[{{Verbiest} {et~al.}(2008){Verbiest}, {Bailes}, {van Straten},
  {Hobbs}, {Edwards}, {Manchester}, {Bhat}, {Sarkissian}, {Jacoby}, \&
  {Kulkarni}}]{vbv+08}
{Verbiest}, J.~P.~W., {Bailes}, M., {van Straten}, W., {Hobbs}, G.~B.,
  {Edwards}, R.~T., {Manchester}, R.~N., {Bhat}, N.~D.~R., {Sarkissian}, J.~M.,
  {Jacoby}, B.~A., \& {Kulkarni}, S.~R. 2008, ApJ, 679, 675

\bibitem[{{Weisberg} {et~al.}(2010){Weisberg}, {Nice}, \& {Taylor}}]{wnt10}
{Weisberg}, J.~M., {Nice}, D.~J., \& {Taylor}, J.~H. 2010, ApJ, 722, 1030

\bibitem[{Weisberg {et~al.}(1989)Weisberg, Romani, \& Taylor}]{wrt89}
Weisberg, J.~M., Romani, R.~W., \& Taylor, J.~H. 1989, ApJ, 347, 1030

\bibitem[{Wolszczan(1991)}]{wol91a}
Wolszczan, A. 1991, Nature, 350, 688

\end{thebibliography}
\bibliographystyle{apjx}
\end{document}